\documentclass[aip,graphicx]{revtex4-1}
\usepackage[utf8]{inputenc}
\pdfoutput=1
\draft % marks overfull lines with a black rule on the right
\usepackage{graphicx}% Include figure files
\usepackage{textcomp}
\usepackage{braket}
\usepackage{amsfonts}
\usepackage{amssymb}
\usepackage{caption}
\usepackage{placeins}
\usepackage{afterpage}
\usepackage{tabularx}

\begin{document}

%\preprint{}

\title{Bad vibrations: Quantum tunnelling in the context of SARS-CoV-2 infection} %Title of paper
\author{Betony Adams}
\affiliation{Quantum Research Group, School of Chemistry and Physics, and National Institute for Theoretical and Computational Sciences, University of KwaZulu-Natal, Durban, 4001, South Africa.}
\affiliation{The Guy Foundation}
\author{Ilya Sinayskiy}
\affiliation{Quantum Research Group, School of Chemistry and Physics, and National Institute for Theoretical and Computational Sciences, University of KwaZulu-Natal, Durban, 4001, South Africa.}
\author{Rienk van Grondelle}%
\affiliation{Department of Biophysics, Faculty of Sciences, VU University Amsterdam, De Boelelaan 1081, 1081 HV, Amsterdam, The Netherlands}
\author{Francesco Petruccione}%
\affiliation{Quantum Research Group, School of Chemistry and Physics, and National Institute for Theoretical and Computational Sciences, University of KwaZulu-Natal, Durban, 4001, South Africa.}
\date{\today}

\begin{abstract}
The SARS-CoV-2 pandemic has added new urgency to the study of viral mechanisms of infection. But while vaccines offer a measure of protection against this specific outbreak, a new era of pandemics has been predicted. In  addition to this, COVID-19 has drawn attention to post-viral syndromes and the healthcare burden they entail. It seems integral that knowledge of viral mechanisms is increased through as wide a research field as possible. To this end we propose that quantum biology might offer essential new insights into the problem, especially with regards to the important first step of virus-host invasion. Research in quantum biology often centres around energy or charge transfer. While this is predominantly in the context of photosynthesis there has also been some suggestion that cellular receptors such as olfactory or neural receptors might employ vibration assisted electron tunnelling to augment the lock-and-key mechanism. Quantum tunnelling has also been observed in enzyme function. Enzymes are implicated in the invasion of host cells by the SARS-CoV-2 virus. Receptors such as olfactory receptors also appear to be disrupted by COVID-19. Building on these observations we investigate the evidence that quantum tunnelling might be important in the context of infection with SARS-CoV-2. We illustrate this with a simple model relating the vibronic mode of, for example, a viral spike protein to the likelihood of charge transfer in an idealised receptor. Our results show a distinct parameter regime in which the vibronic mode of the spike protein enhances electron transfer. With this in mind, novel therapeutics to prevent SARS-CoV-2 transmission could potentially be identified by their vibrational spectra.
\end{abstract}
\maketitle %\maketitle must follow title, authors, abstract and \pacs
\thispagestyle{empty}

\noindent \textbf{Key points:} Quantum biology, lock-and-key, coronavirus, receptors

\noindent \textbf{Key points:} Quantum biology, SARS-CoV-2, lock-and-key,  receptors, vibration assisted tunnelling

\begin{figure}
	\includegraphics[width=\textwidth]{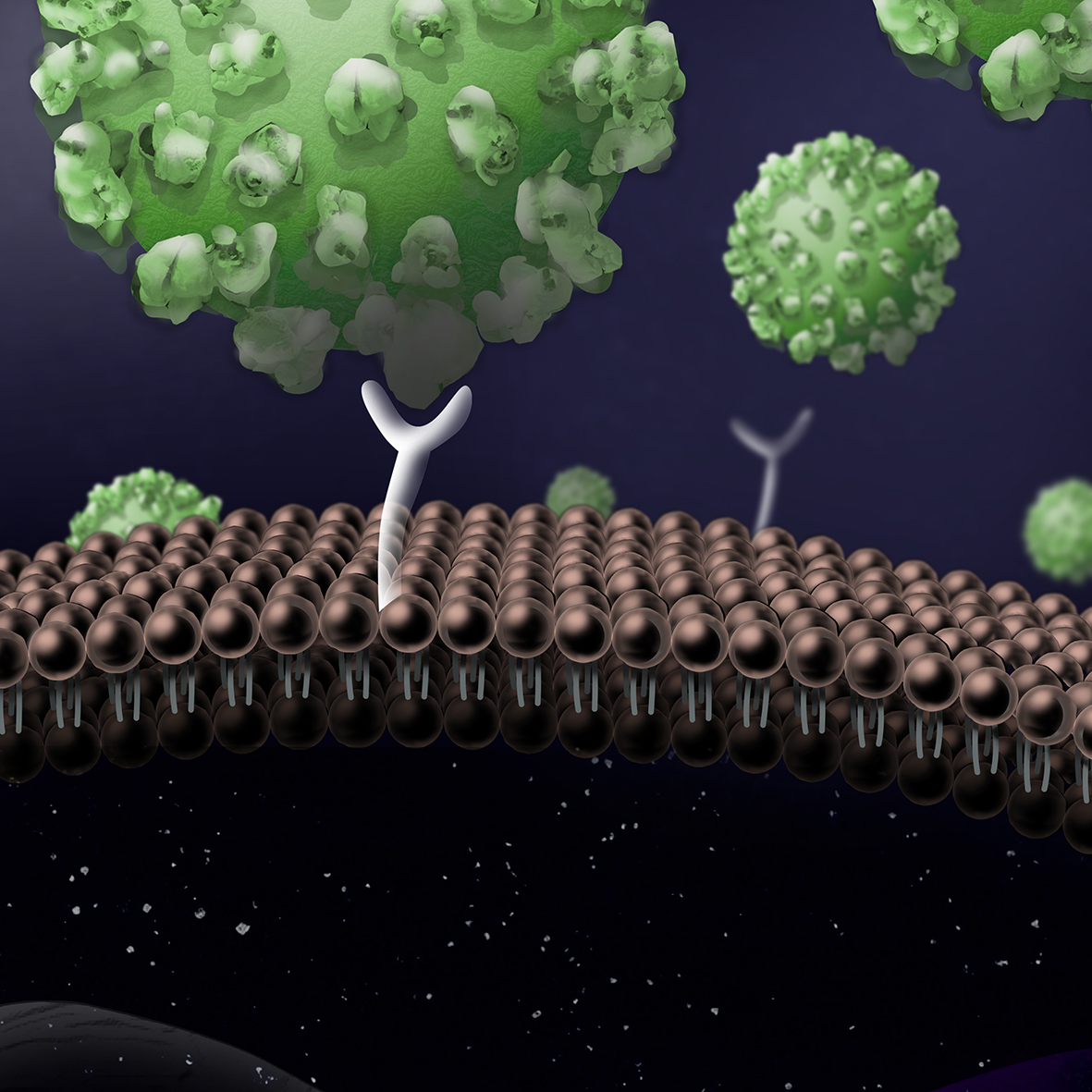}
	\caption{The SARS-CoV-2 spike protein facilitates host cell invasion by binding with cell membrane embedded ACE2 receptors.}
	\label{fig:cell1}
\end{figure}
\section*{Introduction}
Quantum biology is almost as old as quantum mechanics itself \cite{mcfadden}. Where quantum mechanics is often preoccupied with the interaction of light and matter, photosynthesis, that backbone of biology, is the interaction of light with living matter. Bohr himself delivered a lecture which he titled `Light and Life' \cite{bohr}. Schr\"odinger, for his part, wrote \textit{What is Life? The Physical Aspect of the Living Cell}, which served as inspiration for the discovery of DNA \cite{schrodinger,mcfadden}. In a discussion of life, viruses offer a novel case study, straddling, as they do, the properties of both living and non-living systems \cite{hegde}. As such, they might also offer a novel application for quantum biology. There has been some research into investigating how quantum mechanics and viruses intersect. For example, Park \textit{et al}. engineered a virus to obtain enhanced energy transport in excitonic networks \cite{park}. Quantum dots have been used to label viral proteins in an attempt to enhance live imaging of virus-host interactions \cite{yamauchi,qin}. They have also been suggested to have antiviral properties \cite{loczechin,manivannan,kotta}. There has even been an attempt to simulate the life cycle of a virus using quantum gates \cite{shojaie}. Meanwhile, the new coronavirus SARS-CoV-2 has fundamentally changed the world we find ourselves in. While SARS-CoV-2 vaccine development has been integral, there is some suggestion that we have entered a new era of pandemics \cite{peters}. It seems imperative that research into understanding viral mechanisms is accelerated. One way in which this might be achieved is to look at how research outside of established disciplines might allow new insights into physiological mechanisms. Quantum biology, which looks at how non-trivial quantum effects, such as coherence, tunnelling and entanglement, might play a role in biological systems, is one such field of research \cite{marais,alkhalili,mohseni}. The phenomenon of charge transfer is central to much of the research emerging in this field. There is some evidence that coherent energy and charge transfer play a role in photosynthetic networks \cite{marais,engel,brixner,vangrondelle}. This is perhaps the best known application of quantum effects in the biological context. Charge transfer, however, is also important in the context of tunnelling in enzymes, which was first observed a number of decades ago \cite{devault}. Electron tunnelling has also been proposed to be the mechanism, or one of the mechanisms, by which olfactants activate olfactory receptors \cite{turinVTO,reese}. Olfactory receptors are a class of rhodopsin-like receptors known as G-protein coupled receptors, or GPCRs \cite{rosenbaum,gehrckens}. These receptors are implicated in numerous important physiological phenomena from the regulation of inflammation to the binding of neurotransmitters, the latter of which is currently emerging as a new application of vibration-assisted electron tunnelling \cite{hoehn,chee,oh}. Electron tunnelling in these contexts has been investigated as an alternative to the lock-and-key mechanism, a shape-based explanation of receptor binding. Our primary motivation in this paper is the following: given the importance of receptor recognition, binding and activation in the biological context and given the expansion of the lock-and-key model to include the possibility of vibration-assisted tunnelling, it is salient to review the evidence for tunnelling in the context of viral mechanisms of host invasion. This paper follows two main threads. First we outline the case for quantum tunnelling as an alternative or augmentation to the lock-and-key mechanism in the context of enzyme function, olfaction and neurotransmitter reception. We then review those aspects of SARS-CoV-2 infection that suggest a role for quantum tunnelling, specifically the involvement of enzymes as well as certain types of receptors. We also address the possible consequences of this connection using three aspects of SARS-CoV-2 infection: host cell invasion, medical intervention and post-viral syndrome.

\section*{A quantum approach to biological receptor mechanisms}

\subsection*{Broader biological context}
While quantum coherence in photosynthesis might garner more attention, quantum tunnelling in a biological context is arguably the oldest exemplar of the field of quantum biology. In the 1960s, L\"{o}wdin suggested that proton tunnelling might be the physical basis of DNA mutations \cite{mcfadden,lowdin} an idea that is still very much of interest today \cite{slocombe}. Also in the 1960s, enzymes, first described over a century ago by Fischer as operating through a lock-and-key mechanism \cite{fischer}, were subsequently suggested to exploit quantum tunnelling \cite{mcfadden,devault,lowdin}. Both electron and proton tunnelling in enzymes is now a well established field of research \cite{sutcliffe,pusuluk,bothma,moser}. Other biological phenomena have also been characterised as utilising the lock-and-key mechanism of receptor binding. Cellular receptors bind with specific ligands and are integral to signalling processes throughout the body \cite{tripathi}. Receptor recognition and binding accounts for a range of physiological phenomena \cite{uings}. An important class of these receptors are G-protein coupled receptors (GPCRs), examples of which are receptors that mediate the sense of smell or the binding of neurotransmitters in order to open ion channels. What is interesting about GPCRs, in the context of quantum biology, is that they are related to rhodopsin \cite{rosenbaum,gehrckens}. Rhodopsin is a retinal photoreceptor protein which consists of the light-sensitive chromophore retinal in an opsin protein \cite{filipek}. Chromophores are a central theme with respect to quantum effects in photosynthesis, where it has been suggested that quantum coherence might play a role in energy and charge transfer \cite{engel,brixner,vangrondelle}. This is perhaps less a unique feature of photosynthesis than it is due to the more general arrangement of chromophores in a protein \cite{toole}. Chromophores, then, would appear to be important to redox activity in biological materials. There is also a growing focus on the role that the protein scaffold might play in enhancing energy or charge transfer. Far from the warm, wet, decoherent environment that is often cited as an argument against quantum effects in biology, the vibrational or spin states of proteins might be coupled to electronic states in a favourable way \cite{valkunas,marais2}. Interaction with proteins can fundamentally change the properties of a chromophore. Rhodopsin and related opsins, for instance, absorb across a range of frequencies even though they share the same chromophore: retinal. What differs is the opsin protein, which tunes the chromophore's absorption frequency \cite{wang}. This coupling of vibrational to electronic states is most often imagined with respect to the proteins in which the chromophore is embedded, but it might be re-imagined in terms of protein-receptor bonding. While it is still debatable that GPCRs, being related to rhodopsins, operate through a mechanism related to electron transfer, both olfaction and neurotransmitter binding have been of interest in the context of quantum biology. Olfaction has conventionally been described as operating through the recognisable shapes of olfactants \cite{reese}. However, this model has to some extent failed at fully describing the intricacies of our olfactory apparatus. This has given rise to an alternative vibrational theory of olfaction. In this theory, the vibrational spectrum of a ligand rather than its shape is responsible for receptor activation by facilitating electron tunnelling \cite{turinVTO,maniati}. While there is some evidence to support the differentiation of deuterated odorants in various species \cite{turinFF,hoehn,hara,havens}, there remains some scepticism with regards to the theory \cite{block}. The suggestion has been made that the mechanism of olfaction might be closer to a swipe card model, with various different factors contributing \cite{brookes}. More recently, attempts have been made to apply the vibrational theory of olfaction in a different physiological context: the binding of neurotransmitters. Neurotransmitters are integral to the process of neural signal transmission. Signals travelling along the axon of a nerve cell are communicated to adjacent nerve cells by the release of neurotransmitters across the synaptic cleft between cells \cite{lodish}. These neurotransmitters bind to membrane receptors which facilitate the opening of ion channels and thus initiate the activation of nerve cells \cite{lodish}. Theoretical research suggests that the action of specific neurotransmitters such as serotonin and related ligands is correlated to their vibrational spectra \cite{hoehn,hoehn2}. Similar theoretical effects have been suggested for the binding of histamine \cite{oh} and adenosine \cite{chee,chee2} although experimental verification is still lacking. What we would like to highlight, however, is the fact that charge transfer is a well established topic in quantum biology. More specifically, the biological context of this transfer is very often that of membrane-embedded proteins. Within this research much attention is paid to how the biological environment might assist transfer processes, through, for example the vibrations of the protein scaffold or the vibrations of a binding ligand. It is thus potentially informative to consider this in the context of infection by the SARS-CoV-2 virus, which utilises membrane-embedded proteins to invade host cells.

\subsection*{The specific context of SARS-CoV-2}
Our current knowledge of the SARS-CoV-2 virus touches on a number of the specific biological instances outlined in our discussion of tunnelling effects: enzymes, receptor binding and olfaction. Before it can proliferate, the virus first needs to invade its host cell. Research suggests that the SARS-CoV-2 virus most likely invades host cells by interaction with host enzymes, in particular angiotensin converting enzyme (ACE2) \cite{ni,bhalla}. The spike protein of the virus, which is also the target of the vaccine, binds with membrane-embedded ACE2 and facilitates the fusion of virus and host membrane \cite{tang,malik}. In its ordinary cellular context, ACE2 is an enzyme that modulates the form of the GPCR-binding ligand angiotensin, a hormone that is part of the renin-angiotensin-aldosterone system (RAAS). Among other things, angiotensin is important to the balance of vasodilation and vasoconstriction and is integral to cardiovascular function \cite{tikellis,burrell,gheblawi}. The exact interaction of the coronavirus spike protein with its host cell is also of potential importance in light of the new mutated versions of the virus \cite{tulio1,tulio2}. Mutations in the viral genome that code for the spike protein have led to fears of increased transmissability \cite{volz,zhang}. While the ACE2 enzyme is currently the focus of much of the research, other enzymes have also been implicated in SARS-CoV-2 viral infection. Research suggests that the host cell enzyme serine protease TMPRSS2 is necessary for protein priming of the spike protein and facilitates the virus entering the host cell \cite{seth,hoffmann}. Another enzyme, cathepsin L, has been linked to spike protein cleavage and enhanced viral entry into host cells \cite{seth,gomes}. Given the importance of enzymes in viral activity and given the fact that quantum tunnelling plays a role in enzyme activity it seems a closer look at quantum effects in the context of viruses might prove fruitful. More generally, a closer look at receptor mechanisms might offer some insights. Criticism of the vibration assisted tunnelling theory of olfaction often points to the fact that there isn't evidence for electron transfer in olfactory receptors, which are GPCRs \cite{turin2021}. However, a recent, as yet to be reviewed, paper suggests that there is the potential for electron transfer in certain types of GPCRs. Various possibilities for the specific site of this electron transfer are explored in detail in the paper \cite{turin2021}. In the context of SARS-CoV-2, one of these is of particular interest: the disulfide bridge. ACE2 is not a G-protein coupled receptor. Evidence does suggest, however, that its interaction with the spike protein might involve redox activity \cite{singh,keber,hati}. This redox status is also suggested to involve a disulfide bridge. Both spike protein and ACE2 are rich in cysteine residues, which are implicated in intramolecular disulfide bonds \cite{singh}. Indeed, the infectivity of SARS-CoV-2 appears to depend on the disulfide redox potential with resistant animals lacking a redox-active disulfide \cite{singh,keber}. Binding affinity has also been demonstrated to be significantly impaired when the disulfide bonds of ACE2 and SARS-CoV-2 spike proteins are reduced \cite{hati}. ACE2 receptors have a disulfide bridge in common with certain types of GPCRs. But GPCRs themselves also appear to play a role in the disease associated with SARS-CoV-2 infection. The effects of COVID-19 on olfaction have been widely documented as one of the defining symptoms of the disease \cite{lechien,sedaghat}. ACE2 has elevated expression in the olfactory epithelium, where olfactory GPCRs are also located, which might account for the anosmia or altered sense of smell associated with COVID-19 \cite{chen}. GPCRs are also important in the context of COVID-19 inflammatory responses. Increased morbidity has been linked to the cytokine storm induced by the virus. Cytokines are small proteins produced by immune cells. The overproduction and dysregulation of cytokines, however, may lead to tissue damage and death \cite{ragab,hojyo,tisoncik}. As such, cytokines have been suggested as a possible therapeutic target to ameliorate COVID-19 mortality \cite{ragab,hojyo}. A specific class of cytokines known as chemokines, and the chemokine receptor system, have been implicated in the severe clinical sequelae of COVID-19 \cite{coperchini}. Chemokine receptors belong to a group of rhodopsin-like transmembrane GPCRs similar to those activated by neurotransmitters and odorants \cite{gehrckens,lodowski}. Whether or not quantum effects might be at play in any of these facets of SARS-CoV-2 infection and COVID-19 is debatable. However, molecular recognition and binding in the physiological context is integral to viral infection. As a common factor to enzyme function, receptor binding, olfactory symptoms and immune response, it deserves closer scrutiny through as many lenses as possible.
\begin{figure}
	\centering
	\includegraphics[scale=1.1]{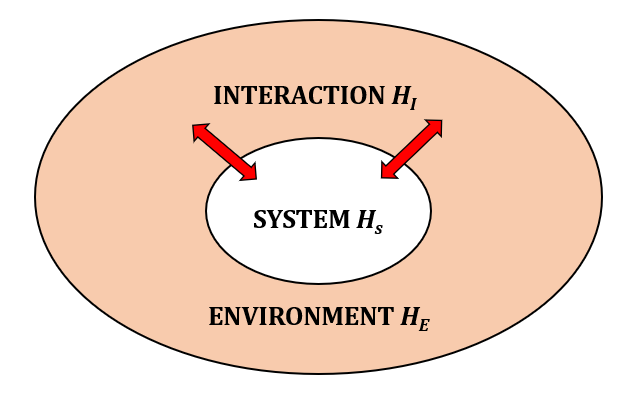}
	\caption{A schematic illustrating the concept of an open quantum system, including the Hamiltonians (H) that mathematically describe the system and environment as well as the interaction between the two. Biological systems interact with their environments and thus are often modelled using an open systems approach. This involves modelling the system and environment as a closed system, the environment is then traced out to arrive at the reduced system dynamics.}
	\label{fig:opensys}
\end{figure}

\subsection*{Relationship between transfer likelihood and vibrational modes}
\begin{figure}
	\centering
	\includegraphics[scale=0.06]{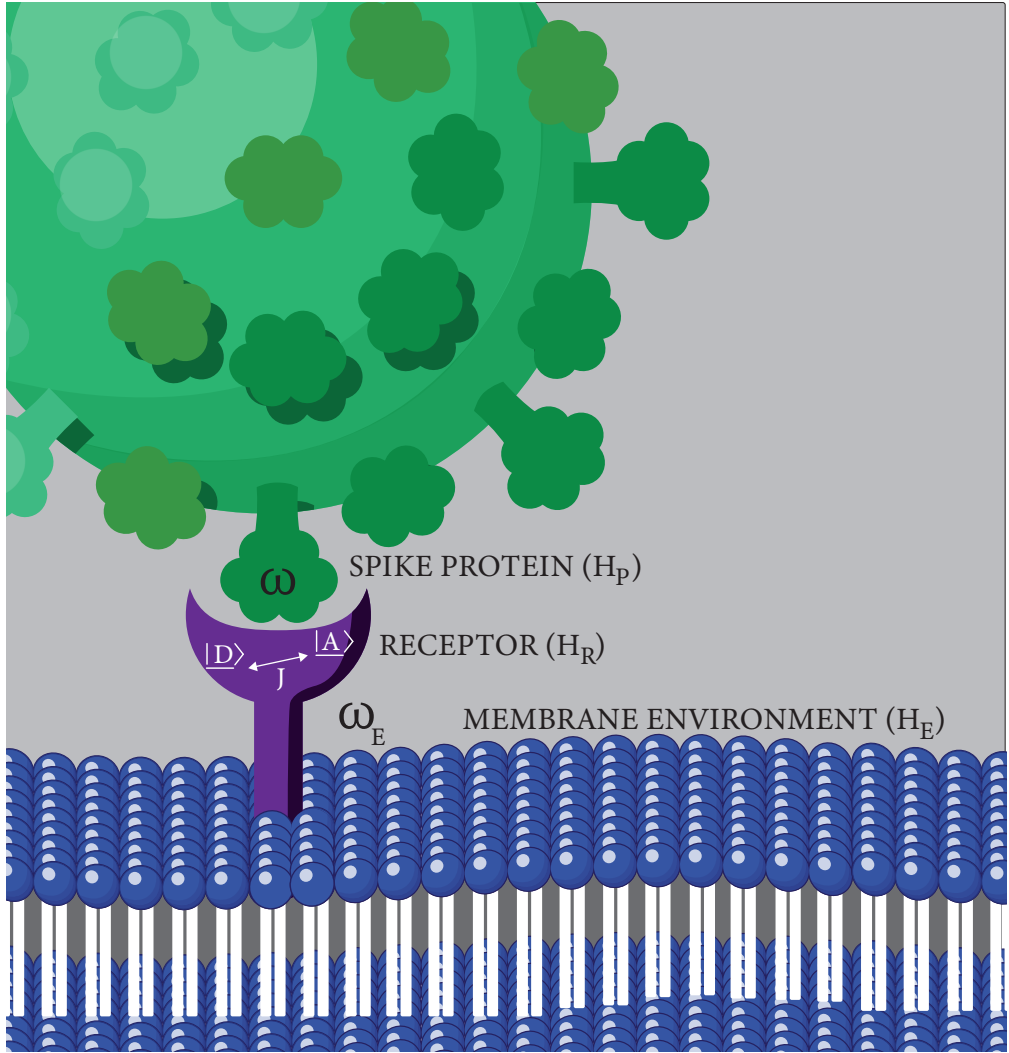}
	\caption{A simplified illustration of vibration assisted tunnelling in the context of SARS-CoV-2 infection. The spike protein vibrational spectrum matches the energy of transition for an electron in the ACE2 receptor, facilitating electron transfer and the activation of the receptor.}
	\label{fig:cell2}
\end{figure}
\subsubsection*{Theory}
We model the interaction of the spike protein and the ACE2 receptor as a vibration assisted electron transfer \cite{marais2,sinayskiy}. Biological systems interact with their environments and thus are often modelled using an open systems approach. In this approach the ligand protein, the receptor and environment are considered a closed system before tracing out the environment to get the reduced dynamics of the system of interest. See Figure 2 for a simple illustration of the concept of an open quantum system, including the Hamiltonians ($H_S$ and $H_E$) that mathematically describe the system and environment as well as the interaction ($H_I$) between the two.  The Hamiltonian describing the system, environment and their interaction is used to derive the master equation \cite{petruccione}. Using this open systems approach and borrowing from a model developed for olfaction \cite{solovyov}, we outline here the relationship between the maximum transfer probability in an ACE2 receptor and its coupling to a single vibronic mode associated with a SARS-CoV-2 spike protein. To simplify things we model the receptor as a dimer (see Figure 3) where the Hamiltonian is given by:
\begin{equation}
H_R=\epsilon_D\ket{D}\bra{D}+\epsilon_A\ket{A}\bra{A}+J(\ket{D}\bra{A}+\ket{A}\bra{D}),
\end{equation}
where $\epsilon_D$ and $\epsilon_A$ are the energy levels of the donor (D) and acceptor (A) levels and $J$ describes the coupling between levels and the likelihood of transition. For a dimer isolated from external interaction, the maximum probability of a transition from donor to acceptor is given by 
\begin{equation}
\textrm{Max}[P_{D\rightarrow A}(t)]=\frac{J^2}{J^2+\Delta ^2},
\end{equation}
where $\Delta=(\epsilon_A-\epsilon_D)/2$. When the energy of the donor and the acceptor are equal then the probability of transfer $[P_{D\rightarrow A}(t_0)]=1$ at time $t_0=\pi/2J$ \cite{sinayskiy}. For a dimer that is not isolated the total Hamiltonian is given by:
\begin{equation}
%H=H_R+H_P+H_E+H_{R-P}+H_{R-E}
H=H_R+H_{R-P}+H_E+H_{R-E}.
\end{equation}
The receptor is represented by the dimer with Hamiltonian, $H_R$. The ligand, in this case the spike protein, is represented as a harmonic oscillator with frequencies associated with the spike protein. This has been incorporated into $H_{R-P}$, the interaction between the receptor and the protein, given by:
\begin{equation}
H_{R-P}=\sum_{i=D,A} \hbar \omega \ket{i}\bra{i}[\gamma_i(a+a^{\dagger})+(a^{\dagger}a+\frac{1}{2})].
\end{equation}
The sum runs over the interaction of the protein with both the donor and the acceptor; the latter is presumed to be zero in the numerical solution. The coupling strength between ligand protein and receptor is given by $\gamma$ and $a$ and $a^{\dagger}$ are the creation and annihilation operators associated with the spike protein vibrations, with associated frequency $\omega$. The Hamiltonians $H_R$ and $H_{R-P}$ represent the system in our model of SARS-CoV-2 receptor tunnelling. The membrane environment $H_E$ and its interaction with the receptor $H_{R-E}$ is similarly approximated as:
\begin{equation}
H_{R-E}=\sum_{i=D,A}\sum_E \hbar \omega_E \ket{i}\bra{i}[\gamma_{iE}(a_E+a_E^{\dagger})+(a_E^{\dagger}a_E+\frac{1}{2})],
\end{equation}
where $\gamma_E$ represents the coupling between the receptor and its membrane environment. This coupling is taken to be weaker than the spike protein coupling to the receptor. The Hamiltonians $H_{R-P}$ and $H_{R-E}$ represent competing interactions, $H_{R-P}$ is essential for receptor recognition of the spike protein, $H_{R-E}$ on the other hand describes undifferentiated coupling to a
multitude of environmental vibronic modes in the vicinity of the receptor. The spike protein vibrations are discriminated from the environment vibrations through their specific frequencies as well as the stronger coupling constants \cite{solovyov}. Here the sum runs over both the donor and the acceptor as well as all the possible environmental vibrations.\\
\\
From the full interaction Hamiltonian we derive a master equation, tracing out the environment and making the standard Born and Markov approximations. This yields:
\begin{equation}
\frac{d}{dt}\rho=-i[H_R+H_{R-P},\rho]+\sum_{j=A,D}\kappa_j (\ket{j}\bra{j}\rho\ket{j}\bra{j}-\frac{1}{2}\{\ket{j}\bra{j},\rho\}),
\end{equation}
where $\rho$ is the reduced density matrix describing the dynamics of the system, $\kappa_D$ and $\kappa_A$ denote respective coupling of donor and acceptor to the external environments. We assume that the initial state of the dimer is at the
donor site, and the vibronic mode is initially in the thermal state (T = 300K).
Using numerical simulation, we would like to understand if coupling
to a vibronic mode could enhance the transition probability in the dimer system:
from donor to acceptor. To this end, we will measure the usefulness of the
vibronic mode coupling by considering the following difference:
\begin{equation}
\Delta P_{\textrm{Max}}=\textrm{Max}[P_{D\rightarrow A}(t)]_{\textrm{vibronic mode}}- \textrm{Max}[P_{D\rightarrow A}(t)],
\end{equation}
where $\textrm{Max}[P_{D\rightarrow A}(t)]_{\textrm{vibronic mode}}$ denotes the maximum probability of the transition in the dimer system calculated from the numerical integration of Equation (6), while $\textrm{Max}[P_{D\rightarrow A}(t)]$ represents the maximum probability of transition
without vibronic mode coupling, given by Equation (2).
\subsubsection*{Results}
\begin{table}[h]
		\caption{Parameters for numerical solution \cite{solovyov,huang}}\label{tab1}%
		\begin{tabular}{@{}lllllll@{}}
			\hline
			& $\epsilon_A-\epsilon_D$ &   J & $\gamma_D$  & $\omega_1$  & $\omega_2$ & $\omega_3$\\
			\hline
			Parameter ranges   & $~ 500-1700 \textrm{cm}^{-1}$  & $0.0001-0.1$ eV  & $0-0.419$ eV & $1690  \textrm{cm}^{-1}$ & $1300  \textrm{cm}^{-1}$ & $1000  \textrm{cm}^{-1}$  \\
			\hline
		\end{tabular}
\end{table}

The difference $\textrm{Max}[P_{D\rightarrow A}(t)]_{\textrm{vibronic mode}}- \textrm{Max}[P_{D\rightarrow A}(t)]$ is plotted for a range of parameters. These parameters have been estimated with respect to comparable biological contexts, in particular Solov'yov \textit{et al.}'s model for vibration-assisted tunnelling in olfactory receptors \cite{solovyov}. The coupling between spike protein and receptor ($\gamma$) is plotted from weak to strong coupling and we have assumed that the coupling only occurs between the spike protein and the donor level, that is $\gamma_A=0$. This coupling strength is plotted as a fraction of the vibronic frequency and hence is unitless. Average energy levels and level coupling are estimated with respect to redox processes in other biological systems \cite{solovyov,fassioli}. This coupling between levels in the dimer (J) is then plotted for a range of different values for each vibronic frequency: 0.0001, 0.001, 0.01 and 0.1 eV. The frequencies of the vibronic mode of the spike protein are taken from studies that investigate SARS-CoV-2 using Raman spectroscopy \cite{huang}. This includes the vibrational spectrum for both the S1 and S2 subunits as well as the full spike protein and the receptor binding domain \cite{huang}.  We have plotted results for three different frequencies $\omega_1 = 0.2095$ eV ($1690  \textrm{cm}^{-1}$), $\omega_2 = 0.1612$ eV ($1300  \textrm{cm}^{-1}$) and $\omega_3 = 0.1240$ eV ($1000  \textrm{cm}^{-1}$), see Figures 4--6. All three cases show a parameter regime in which the vibronic mode enhances electron transfer probability. White regions represent parameter regions where vibronic modes do not enhance electron transfer probability. Redder regions are where vibronic modes have a negative effect on electron transfer probability. Bluer regions represent where vibronic modes enhance electron transfer probability. Vibronic modes have a marked effect on transfer probability in a selective parameter regime, with this effect growing as coupling strength between levels increases. However when coupling between dimer levels is too weak (J = 0.0001 eV) the vibronic mode has no effect. And when the coupling is too strong (J = 0.1 eV) the vibronic mode begins to have a negative effect. This suggests a distinct biologically relevant parameter window in which vibration-assisted tunnelling takes place. Although the different frequencies of the vibronic mode display similar effects, the higher frequency (see Figure 4) appears to only have a single parameter regime in which the vibronic mode shows marked enhancement. However, lower frequencies (see Figures 5 and 6) appear to have two regimes in which the vibronic mode enhances transfer probability. Whereas Figures 4, 5 and 6 demonstrate the effects of the proposed model over a range of biologically viable parameters, we were also interested in testing the model at the extremes of these parameters.	To this end we have chosen the dimer coupling strength with the least favourable outcome (J = 0.1 eV) and plotted the effects of the vibronic coupling ($\gamma_D$) for very strong coupling to the vibronic mode. While this has an unfavourable effect at small dimer detuning, at larger dimer detuning the effect becomes favourable, showing enhanced transfer probability in distinct regions, especially for higher frequencies of the vibronic mode (see Figure 7). 				
\begin{figure*}[t!]
	\centering
	\includegraphics[scale=0.55]{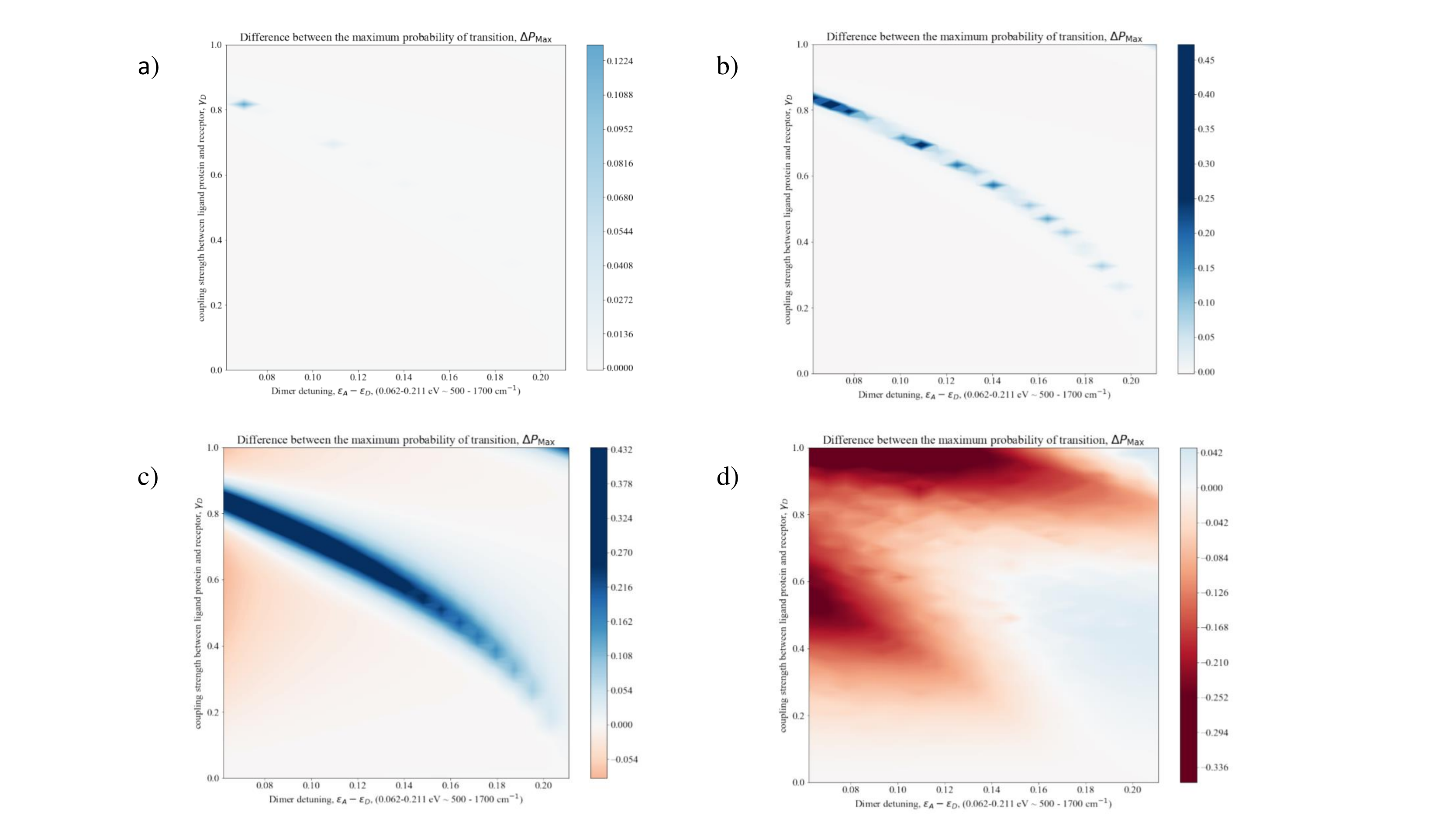}
	\caption{$\Delta P_{\textrm{Max}}$ as a function of the dimer detuning and the coupling strength between donor and vibronic mode. Results show coupling to vibronic mode frequency $\omega = 0.2095$ eV ($1690  \textrm{cm}^{-1}$). Bluer regions show enhanced transfer with vibronic modes, white regions show no enhancement while redder regions demonstrate decreased transfer. Graphs (a)--(d) show the effects of increasing dimer coupling strength by an order of magnitude from J = 0.0001 eV to J = 0.1 eV. The results effectively illustrate the window of (biologically relevant) parameters within which vibration-assisted tunnelling has an effect.} 
\end{figure*}

\begin{figure*}[t!]
	\centering
	\includegraphics[scale=0.55]{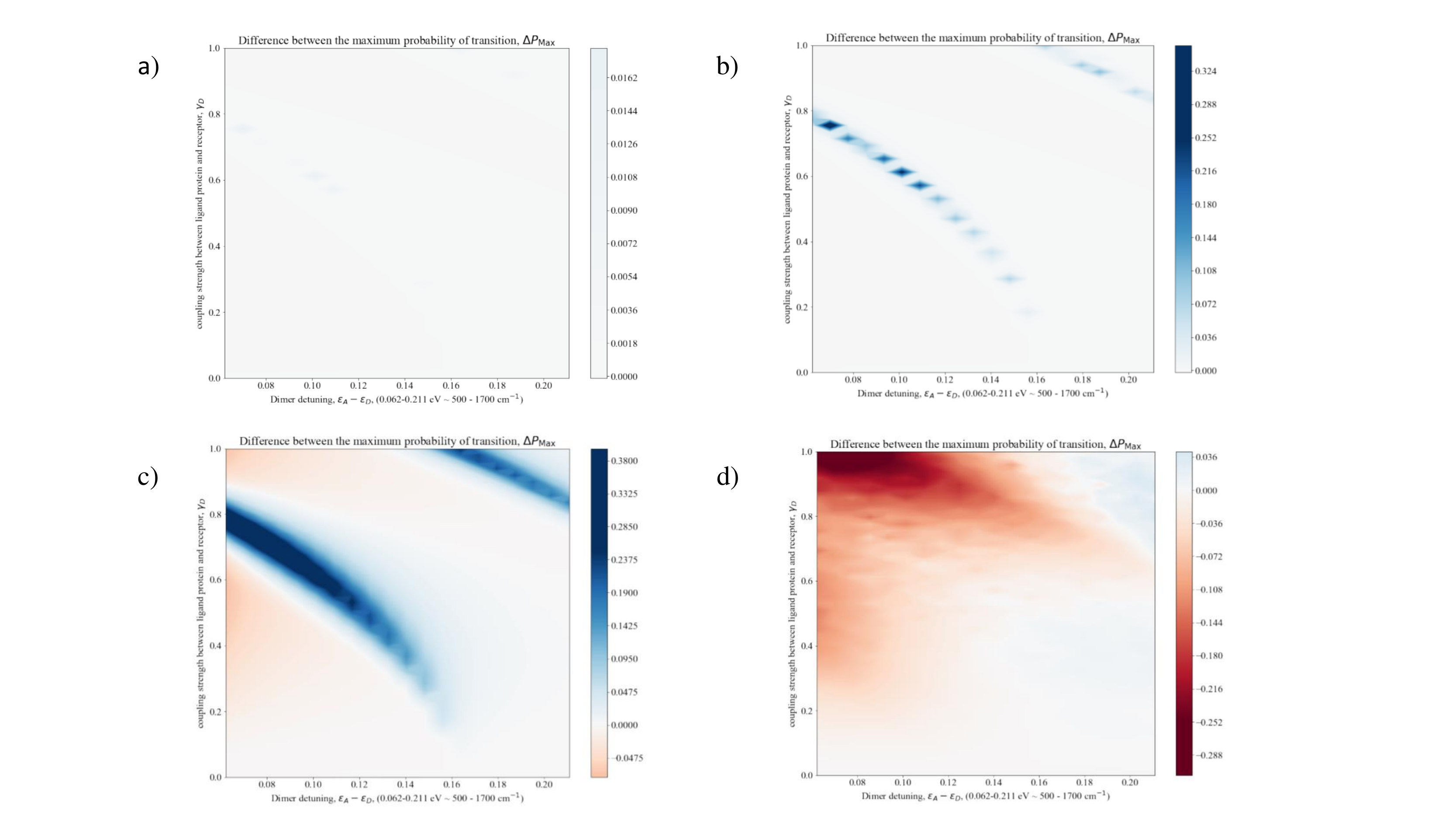}
	\caption{$\Delta P_{\textrm{Max}}$ as a function of the dimer detuning and the coupling strength between donor and vibronic mode. Results show coupling to vibronic mode frequency $\omega = 0.1612$ eV ($1300  \textrm{cm}^{-1}$). Bluer regions show enhanced transfer with vibronic modes, white regions show no enhancement while redder regions demonstrate decreased transfer. Graphs (a)--(d) show the effects of increasing dimer coupling strength by an order of magnitude from J = 0.0001 eV to J = 0.1 eV. The results effectively illustrate the window of (biologically relevant) parameters within which vibration-assisted tunnelling has an effect.} 
\end{figure*}

\begin{figure*}[t!]
	\centering
	\includegraphics[scale=0.55]{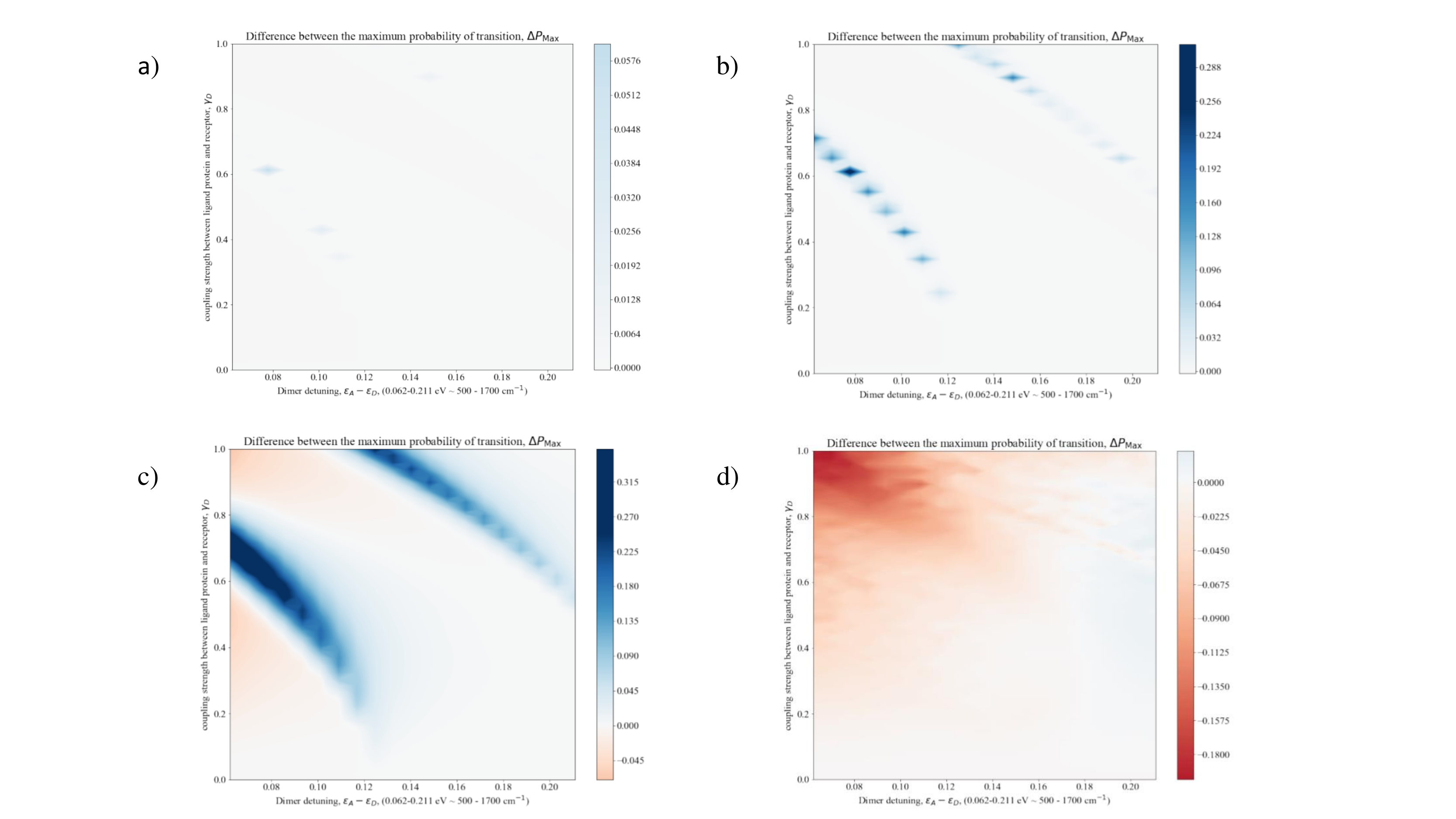}
	\caption{$\Delta P_{\textrm{Max}}$ as a function of the dimer detuning and the coupling strength between donor and vibronic mode. Results show coupling to vibronic mode frequency $\omega = 0.1240$ eV ($1000  \textrm{cm}^{-1}$). Bluer regions show enhanced transfer with vibronic modes, white regions show no enhancement while redder regions demonstrate decreased transfer. Graphs (a)--(d) show the effects of increasing dimer coupling strength by an order of magnitude from J = 0.0001 eV to J = 0.1 eV. The results effectively illustrate the window of (biologically relevant) parameters within which vibration-assisted tunnelling has an effect.} 
\end{figure*}

\begin{figure*}[t!]
	\centering
	\includegraphics[scale=0.5]{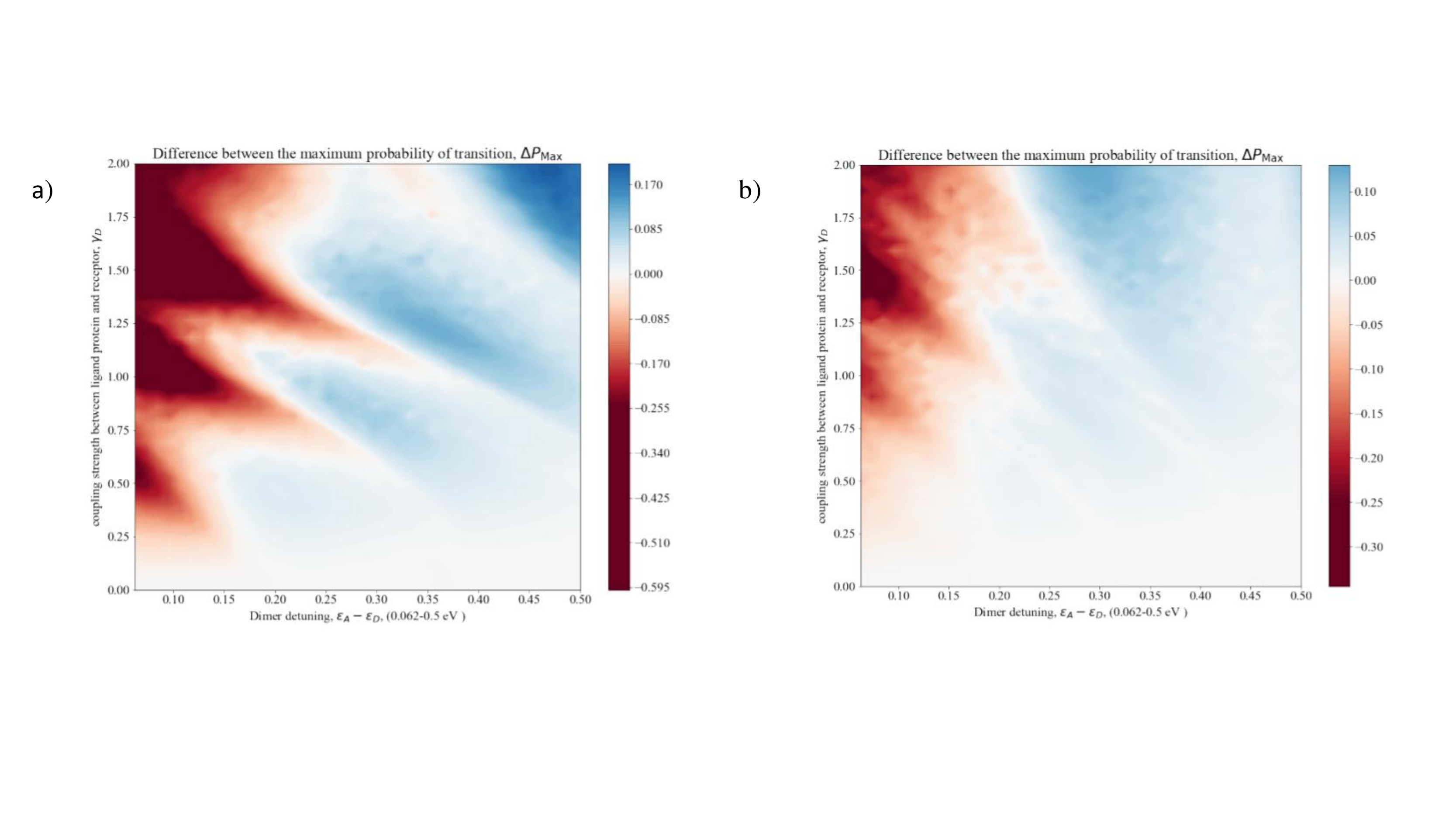}
	\caption{Comparison of highest and lowest vibronic frequencies $\omega = 0.2095$ eV ($1690  \textrm{cm}^{-1}$) and $\omega = 0.1240$ eV ($1000  \textrm{cm}^{-1}$) for very strong coupling to vibronic mode. Coupling strength with respect to frequency is plotted up to a maximum of 2 rather than 1, as in previous figures. The detuning between the dimer levels is plotted up to 0.5 eV rather than 0.2 eV as in previous figures.} 
\end{figure*}								
\subsection*{Possible implications}
\subsubsection*{Novel therapies for COVID-19}
A better understanding of the various ways in which viruses and host cells interact through molecular recognition and binding might also lead to novel treatments for COVID-19. It has already been suggested that treatment with ACE2 inhibitors might have an effect on the severity of the disease. However, reviews are mixed as to whether this treatment may help or harm \cite{cox,nunes}. It has also been suggested that introducing soluble ACE2 might work against the virus by binding to the viral spike protein before it can bind to membrane ACE2 receptors \cite{aziz}. ACE2 receptors also catalyse the different forms of angiotensin that bind with GPC receptors, in particular angiotensin receptors \cite{tikellis,burrell}. Whereas ACE2 inhibitors prevent the production of different angiotensin proteins, angiotensin receptor blockers prevent the action of angiotensin proteins. What is interesting is that there is some evidence that targeting angiotensin receptors with receptor blockers also confers some protection against the SARS-CoV-2 virus \cite{yang2,derington,pirola}. Mechanisms of receptor binding are an important factor in pharmaceutical development. GPCRs, for instance, are one of the major targets of many pharmaceutical drugs and bind to a broad spectrum of ligands \cite{sriram}. The specificity of this binding is complicated by receptor promiscuity and related antagonism or partial agonism \cite{berg}. Does the spike protein's affinity for ACE2 receptors also mean an affinity or partial agony for related angiotensin receptors, and GPCRs more generally? The involvement of GPCRs in viral infection has already been investigated in the context of the Ebola and Marburg viruses which employ a glycoprotein to facilitate host invasion. Chemical screening allowed for the indentification of a number of inhibitory agonists of various GPCRs, including receptors for serotonin, histamine, and acetylcholine, which showed antiviral action \cite{cheng}. The SARS-CoV-2 spike protein is also a glycoprotein and screening for appropriate GPCR agonists might also yield novel therapeutic options. Tryptophans, for instance, have been demonstrated to play a potentially therapeutic role in SARS-CoV-2 infection \cite{essa,anderson}. Tryptophan is the precursor to serotonin, perhaps most widely known as a neurotransmitter implicated in mental illnesses such as depression \cite{delgado}. In the field of quantum biology there has been recent interest in whether quantum effects play a role in the binding of serotonin to its relevant GPCRs \cite{hoehn,hoehn2}. What makes this more interesting in the context of COVID-19 is that antidepressants known as selective serotonin reuptake inhibitors (SSRIs) have been shown to be somewhat effective against  the SARS-CoV-2 virus \cite{lenze,hoertel,hamed}. While the studies are far from being conclusive this isn't the first time that these antidepressants have been investigated as antivirals, although the mechanism of action appears to be less to do with serotonin modulation than viral replication \cite{zuo,benkahla}. Recent studies also suggest that SSRIs are not alone in their potential as a COVID-19 therapeutic. Other antidepressants, including venlafaxine, appear to improve the prognosis of patients hospitalised with COVID-19 \cite{hoertel}. Venlafaxine targets both serotonin and norepinephrine, both of which bind to GPCRs to activate ion channels \cite{mccorvy,vasudevan}. Other GPCR agonists have also been reported to have some effect in mitigating COVID-19 infection. Histamine, which plays a role in neuromodulation and transmission in addition to mediating immune and allergy responses, is an agonist that binds to a number of GPCRs \cite{nuutinen}. Antihistamines, on the other hand, bind to histamine GPCRs and act as blockers or reverse agonists. There is some evidence that certain antihistamines protect against SARS-CoV-2 infection by disrupting the way in which the virus binds to its host cell \cite{reznikov}. There is still some doubt as to whether nicotinic receptors act, as least in part, as GPCRs \cite{kabbani}. These receptors, which bind the neurotransmitter acetylcholine as well as the agonist nicotine, do however open ion channels \cite{govind,changeux}. Whereas there is little debate that smoking itself offers any protection against COVID-19, surprising statistics around the hospitalisation of smokers and non-smokers with COVID-19 has led to some speculation that nicotine might be a potentially therapeutic intervention against severe disease. It remains unclear, however, whether the effects of nicotine are helpful or harmful \cite{leung,farsalinos}. The generalisation of the vibrational theory of olfaction to the binding of neurotransmitters has led to some suggestion that the agonist and antagonist action of certain ligands might be classified according to their vibrational spectra \cite{chee,oh,chee2}. In the context of SARS-CoV-2, new therapeutics might be discovered by screening and selecting related ligands through, for example, their Raman spectra. The different vibrational spectra of mutated spike proteins might also allow some prediction of the infectivity of new variants of SARS-CoV-2. This paper has focused on the SARS-CoV-2 spike protein binding to ACE2 receptors and its possible interaction with G-protein coupled receptors. In particular it has looked at at how the spike protein vibronic mode might alter electron transfer in certain receptors. What many of these receptors potentially have in common is a disulfide bridge and the potential for redox activity \cite{turin2021,singh}. Targeting the disulfide bonds has been shown to modulate host cell invasion \cite{keber}. This interest in the redox activity of the SARS-CoV-2 virus might also be extended beyond host cell invasion to offer insights on other aspects of COVID-19. ACE2 is a regulator of oxidative stress and it has been suggested that increased vulnerability to COVID-19 is related to increased oxidative stress, through factors such as increased age or underlying health issues \cite{singh,nunn,chernyak,forcados}. Redox reactions proliferate in the body, not least in the electron transport chains within mitochondria. The spike protein has also been shown to directly modulate mitochondrial activity, most probably through ACE2 signalling \cite{lei}. Whether or not the spike protein is involved, a growing body of research suggests that mitochondria are implicated in COVID-19 and as such might inform novel therapeutic options \cite{nunn,riya,paul}.

.
\subsubsection*{Post-viral syndrome: Long COVID}

While redox considerations, receptor binding mechanisms and the involvement of GPCRs in SARS-CoV-2 infection might lead to possible novel treatments for the disease, it might also offer insights into the post-viral condition referred to as `long COVID' \cite{callard}. Research into this condition is still in the very early stages, and much of it is focused on the urgent need for more research to be undertaken, due to the large number of people who appear to experience long term symptoms relating to COVID-19 \cite{callard,nalbandien,nabavi}. Long COVID is not necessarily correlated with the severity of the active infection, with some patients reporting mild symptoms during the initial, acute stage of the disease before going on to experience lingering sequelae \cite{callard,ladds}. Some of the long-term effects may be due to damage wrought by COVID-19 to organs such as the lungs and heart\cite{marshall,rio,ladds}. However, an appreciable portion of those reporting long term effects show no obvious biomarkers to account for their disorienting collection of symptoms, ranging from fatigue and joint pain to brain fog, memory problems, mood swings and mental illness \cite{marshall,ladds}. In its lack of defining mechanism and broad range of symptoms, long COVID resembles the condition that is sometimes called myalgic encephalomyelitis (ME) or chronic fatigue syndrome (CFS) \cite{callard,marshall}. It is well known that viral infections can cause a range of long term effects \cite{bannister,archer}. Despite this, the condition remains under-researched, and, due to its various psychological manifestations often prompts suggestions that the syndrome is psychosomatic \cite{underhill,friedman}. There is some research, however, that points to the involvement of GPCRs in both ME and CFS, in particular the disruption of GPCR function by autoantibodies \cite{johnston,loebel,wirth}. GPC receptors control a wide range of essential functions and bind to a broad spectrum of different ligands, which makes them an excellent target for drug development. However this also means that specific ligands might interact with receptors other than their primary receptor, contributing, for example, to the side-effects of a GPCR-targeting drug \cite{sriram}. Whether long COVID involves GPCR disruption remains to be seen. But if ACE2 can bind both the SARS-CoV-2 virus as well as molecules such as angiotensin, then perhaps the virus can mimic, at least partially, the way in which angiotensin binds to GPCRs, either through specific viral proteins or through autoantibodies. GPCR disruption would also potentially explain the wide-ranging array of symptoms reported by long COVID sufferers, as GPCRs are implicated in many different physiological processes \cite{rosenbaum}. GPCR involvement in ion channel action might also prove an avenue of research for potential therapeutics. Viruses or the virome play an important, as yet not fully understood, role in the body \cite{adiliaghdam}. It is thus conceivable that long COVID is a manifestation of some aspect of the SARS-CoV-2 virus being incorporated into host cells even beyond the infected stage. Viroporins, for instance are viral proteins that can oligomerise in host cell membranes to form ion channels of their own \cite{nieva,scott}. The physiological mechanisms behind both ME and CFS have been suggested to involve ion channels \cite{chaudhuri}. It might thus be interesting to investigate this in the context of long COVID, particularly as the envelope protein found in the SARS-CoV-2 virus has been shown to have viroporin capabilities \cite{mandala,schoeman}. Ion channels are instrumental in maintaining membrane potential. While it is more common knowledge that membrane potential is integral to the activation and efficient function of nerve cells, all cells have an associated membrane potential. It is also becoming clearer that this membrane potential plays an important role in disease, not least cancer \cite{levin,jolley}. Membrane potential is also integral to mitochondrial function, where it is coupled to energy and charge transfer in metabolic processes. Both ME and CFS have been suggested to involve metabolic processes \cite{nilsson,tomas}.
It has recently been suggested that long COVID resembles ME and CFS in redox imbalance, inflammation, an impaired ability to generate adenosine triphosphate (ATP), and general hypometabolic state, all of which implicate mitochondria in the process \cite{nunn,paul}. There is some evidence that supporting redox processes, through co-enzymes for instance, may help with metabolic illnesses \cite{quinzii,hargreaves}. There is even some evidence that the ingestion of chlorophyll, the chromophore central to photosynthesis, might alter mitochondrial ATP production \cite{ilyas}. It is thus perhaps not too much of a stretch to suggest that elements of the SARS-CoV-2 virus might be incorporated into the redox function of mitochondria. 

\section*{Future Perspectives}
This paper has been structured around two related assertions. The first of these builds on the possibility that the lock-and-key or shape-based mechanism used to describe a number of biological phenomena might be replaced or augmented by a quantum tunnelling mechanism. As such, quantum tunnelling is worth investigating in a variety of contexts where molecular recognition and reception play a role; in particular, in this paper, in the context of membrane-receptor binding of SARS-CoV-2. The second assertion addresses this specific context and the way in which quantum tunnelling might be implicated in the receptor binding of the SARS-CoV-2 spike protein, either through the role of enzymes or the involvement of GPCRs. In the event of the latter, the degeneracy that no doubt allows biological systems their flexibility, also allows for the wide range of symptoms attributed to COVID-19 and long COVID. If GPCR-targeting pharmaceuticals can target more than the specific receptor they are aimed at, causing a variety of side-effects, then perhaps the spike protein behaves in a similar manner. And perhaps a better understanding of receptor recognition might contribute to better medical intervention. Regardless of whether these assertions prove to be true, the point remains that questions of interest in quantum biology, such as tunnelling in the context of enzymes and GPC receptors, intersect with some of the open questions in SARS-CoV-2 research. As such, quantum biology can add to the store of knowledge that will offer protection against the SARS-CoV-2 virus as well as novel future viruses. Techniques used in quantum biology, such as the comparison of vibrational spectra to predict GPCR agonist potency \cite{hoehn}, might also inform approaches to virus research. The vibrational characteristics of the SARS-CoV-2 spike protein have already been used to gain insight into its structure by translating the protein into music \cite{buehler}. More prosaically, SARS-CoV-2 infection has also been investigated using Raman spectroscopy \cite{zhang2,jinglin}. This might be extended to comparing the spectra of mutated spike proteins and whether this correlates with how infectious the mutated versions are \cite{li}. Quantum biology might also offer some insights into the long term debilitating effects of COVID-19 and shape possible treatments. While the focus in this paper has been on quantum tunnelling in enzymes and GPC receptors, other related mechanisms of interest in quantum biology offer further avenues of research. ACE2, for example, is a regulator of oxidative stress \cite{singh}. Reactive oxygen species (ROS) have also been implicated in GPCR activity \cite{fukai}. ROS are important signalling molecules but are also implicated in cellular inflammation and damage \cite{auten}. The production of ROS has been demonstrated to be sensitive to magnetic fields, a fact that has been attributed to the involvement of radical pairs, one of the primary topics of interest in quantum biology \cite{usselman}. Inflammation is a contributing factor in both acute infection with the SARS-CoV-2 virus as well as in long COVID \cite{tay,doykov}. Recently it has been shown that the application of electromagnetic fields can significantly ameliorate the inflammation associated with COVID-19 \cite{ahmad}. Progress made in quantum biology may thus have even more to offer the study of viruses than the preliminary ideas laid out in this paper.

\noindent\textbf{Acknowledgements}\\
B.A, I.S. and F.P. were supported by the South African Research Chair Initiative of the Department of Science and Technology and the National Research Foundation. Thank you to Angela Illing for the diagrams.\\

\noindent\textbf{Competing interests}\\
The authors declare no competing interests.\\

\end{document}